\documentclass[mathleft,fleqn,%
]{an}
\usepackage{graphicx}
\usepackage[varg]{txfonts}
\begin{document}

% The following seven commands are intended for editorial usage and
% should be ignored by the author(s).
\Pagespan{1}{}% Document's page range. 
% If second parameter is left empty, the last page is computed
% automatically.
\Yearpublication{2016}%
\Yearsubmission{2015}%
\Month{0}%   
\Volume{999}%  
\Issue{0}% 
\DOI{asna.201400000}% 

\title{Dynamical and chemical evolution of the thin disc}

\author{
A. Just\thanks{Corresponding author: {just@ari.uni-heidelberg.de}}
\and J. Rybizki\thanks{Fellow of the International Max Planck Research School for Astronomy
and Cosmic Physics at the University of Heidelberg (IMPRS-HD).}\\
}
\titlerunning{Chemodynamical disc evolution}
\authorrunning{Just \& Rybizki}
\institute{
Astronomisches Rechen-Institut / Zentrum f\"ur Astronomie der Universit\"at Heidelberg, 
M\"onchhofstrasse 12-14,
69120 Heidelberg, Germany }

\received{XXXX}
\accepted{XXXX}
\publonline{XXXX}


\abstract{%
Our detailed analytic local disc model (JJ-model) quantifies the interrelation between kinematic properties (e.g. velocity dispersions and asymmetric drift), spatial parameters (scale-lengths and vertical density profiles), and properties of stellar sub-populations (age and abundance distributions). Any consistent radial extension of the disc evolution model should predict specific features in the different distribution functions and in their correlations. Large spectroscopic surveys (SEGUE, RAVE, APOGEE, Gaia-ESO) allow significant constraints on the long-term evolution of the thin disc.
We discuss the qualitative difference of correlations (like the $\alpha$-enhancement as function of metallicity) and distribution functions (e.g. in [Mg/H] or [Fe/H]) for the construction of a disc model. In the framework of the JJ-model we build a local chemical enrichment model and show that significant vertical gradients for main sequence and red clump stars are expected in the thin disc. A Jeans analysis of the asymmetric drift provides a link to the radial structure of the disc. The derived metallicity-dependent radial scale-lengths can be combined in the future with the abundance distributions at different Galactocentric distances to construct full disc models. We expect to be able to constrain possible scenarios of inside-out growth of the thin disc and to characterise those populations, which require significant radial migration.
}

\maketitle

\section{Milky Way disc models}
The most elaborate model of the present day Milky Way is the Besan\c{c}on Galaxy Model (BGM) including extinction, the bar, spiral arms and the warp. The most recent version of the BGM proposes an update of the initial mass function (IMF) and the star formation history (SFR) of the disc (Czekaj et al. 2014). Nevertheless there are still open issues to be solved concerning the degeneracy of the SFR and IMF and a consistent chemical abundance model. Our alternative local disc model (JJ-model) based on the kinematics of main sequence stars (Just \& Jahrei{\ss} 2010), the stellar content in the solar neighbourhood (Rybizki \& Just 2015) and Sloan Digital Sky Survey (SDSS) star counts to the north Galactic pole (Just et al. 2011) has a significantly higher accuracy compared to the old BGM (Gao et al. 2013). In the JJ-model the local SFR, IMF, AVR (age--velocity dispersion relation) are determined self-consistently and it includes a simple chemical enrichment model. In order to extend the JJ-model over the full radial range of the disc, we have used the Jeans equation for the asymmetric drift to connect local dynamics with the radial scale-lengths of stellar sub-populations (Golubov et al. 2013). Based on RAVE (RAdial Velocity Experiment, Kordopatis et al. 2013) data we found an increasing scale-length with decreasing metallicity, which is consistent with a negative overall metallicity gradient of the disc. On the other hand Milky Way-like galaxies show a radial colour gradient of the disc to be bluer and younger in the outer part. Combining both observations immediately shows that the chemical enrichment in the inner disc must be faster/larger compared to the outer disc. There are different paths of this inside-out growth of the disc. Models based on a Kennicutt-Schmidt law grow faster in the inner part of the disc due to higher densities. Alternatively, there may be a delay in star formation in the outer disc induced by the star formation threshold. It is a challenge to disentangle these different scenarios and to build a consistent disc model. Precise ages of main sequence stars (calibrated by astroseismology, e.g.) would be the silver bullet to solve this problem, but currently the more promising support comes from observations of abundances and abundance ratios of heavy elements for large stellar samples distributed over the full disc range. The $\alpha$-enhancement is a good tracer for the enrichment timescale and can be used to infer the evolution history of the disc. It can also be used to disentangle the thin and thick disc without introducing a kinematic bias (see e.g. Lee et al. 2011).

\begin{figure}
\includegraphics[width=0.48\textwidth]{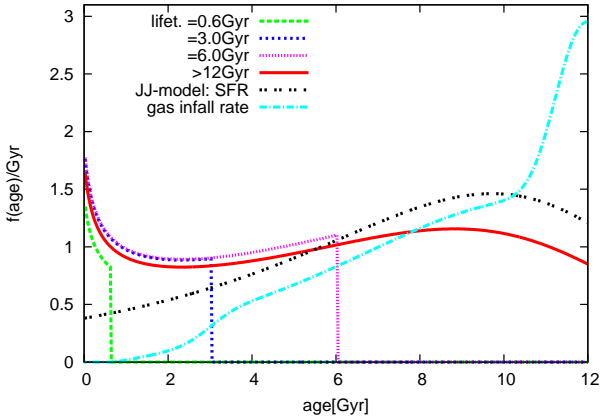} %{ageMS}
\caption{Normalised present day age distributions of main sequence stars with different lifetimes in the solar neighbourhood. For comparison the normalised SFR and the corresponding gas infall rate are shown.}
\label{just-fig1}
\end{figure}

Observational data of stellar populations appear in two fundamentally different kinds leading to very different constraints on the models. Star counts (like luminosity functions, colour-magnitude diagrams, metallicity distribution functions (MDF), or velocity distribution functions) provide quantitative information about the stellar populations. They require complete datasets or a detailed understanding of incompleteness. On the other hand, correlations (like the AVR, the $\alpha$-enhancement as function of metallicity, or the asymmetric drift--metallicity relation) tell us something about the dominating physical processes of the disc evolution. They require an unbiased selection of stars (or a grasp of the biases).

\section{Chemical enrichment}

Many analytic chemical evolution models (see Matteucci 2012 for an overview) are still local, annuli in case of the disc, and rely at least partly on the instantaneous recycling approximation (IRA). In order to overcome these restrictions,  Sch\"onrich \& Binney (2009) quantified the impact of radial mixing of stars and gas in an analytic model, whereas models based on numerical simulations start to incorporate a chemical evolution network to reproduce abundance distributions in detail (e.g. Minchev et al. 2014; Kubryk et al. 2015a,b).

\begin{figure}
\includegraphics[width=0.48\textwidth]{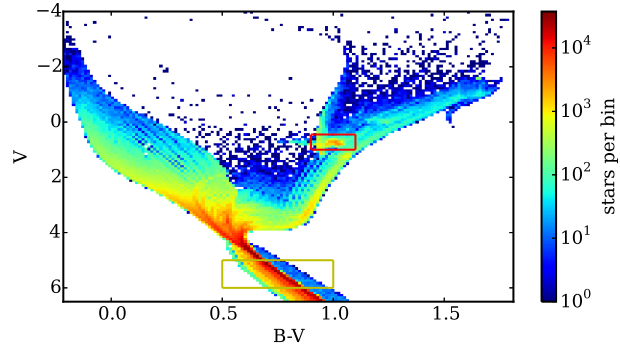} %{selection_with_cbar}
\caption{Model CMD with unweighted isochrones corresponding to a uniform age distribution. Red clump and lower main sequence selection boxes are shown in red and yellow, respectively.}
\label{just-fig2}
\end{figure}

As a first step to a consistent chemical enrichment in the framework of the JJ-model (default model A of Just \& Jahrei{\ss} 2010) we analyse the $\alpha$-enhancement in a local one-zone model with gas infall. For comparison we use two local volume-complete datasets of main sequence stars, namely the Geneva-Copenhagen sample (GCS), where the $\alpha$-enhancement was determined by Str\"omgren photometry (Casagrande 2011), and the Hipparcos sample, where the abundances were determined by high resolution spectra (Fuhrmann 2011). The chemical evolution depends on the SFR, the IMF and the gas infall rate. For comparison with the local sample, the local age distribution is needed additionally, since the dynamical heating given by the AVR results in an age-dependent vertical dilution factor measured by the vertical thickness of the corresponding sub-populations. 
The SFR and the local age distribution for main sequence stars with different lifetimes, normalised to an average of 1/Gyr, are shown in Fig.~\ref{just-fig1} for the JJ-model. 
In order to reproduce the age-selection for specific types of stars we feed the IMF and chemical enrichment of the JJ-model combined with a flat SFR into the {\sc Galaxia} tool (Sharma et al. 2011) with PARSEC isochrones (Bressan et al. 2012) and create a large Mock sample (see Fig.~\ref{just-fig2} for the CMD). Then we select the corresponding region in the CMD and weight the selected stars with the age distribution given by the JJ-model for the volume of interest (e.g. the yellow box in Fig.~\ref{just-fig2} and the red line in Fig.~\ref{just-fig1} to reproduce the local G dwarf sample).

There are two free parameters in the model. The high-mass slope of the IMF above 8\,M$_{\odot}$ was not determined in Rybizki \& Just (2015), because there are no O- and early B-type stars in the solar neighbourhood. Secondly, the gas infall rate can be varied to reproduce the abundance distributions and the $\alpha$-enhancement. The only boundary condition here is the present day surface density of gas. 

Some  $\alpha$-elements (we use oxygen and magnesium) are predominantly produced by core collapse supernovae (SN2) of massive stars. Due to the short lifetime of the progenitor stars, the IRA is a good approximation and we can use the observed O- and Mg-abundance distributions to derive the gas infall rate. The result of our fiducial model (with high-mass slope -2.7 for the IMF (the Salpeter IMF has a slope of -2.35) and SN2 yields of Fran\c{c}ois et al. (2004) is shown in Fig.~\ref{just-fig1} with the same scaling as the SFR.  

\begin{figure}
\includegraphics[width=0.48\textwidth]{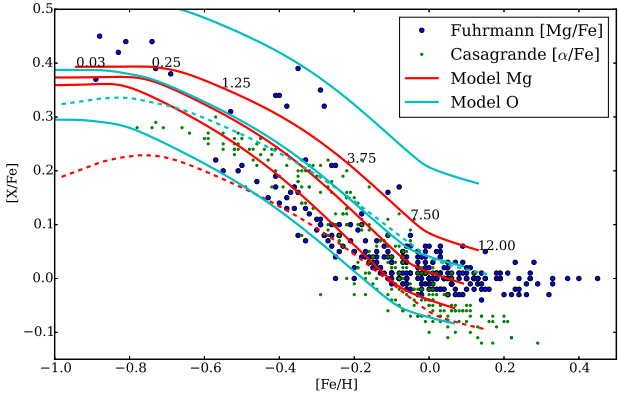} %{high-mass}
\caption{The symbols show the $\alpha$-enhancement for the datasets of Casagrande and Fuhrmann. The full lines (red for Mg, cyan for O) show the Fran\c{c}ois yields combined with different IMF slopes (-2.3, -2.7, -3.0 from top to bottom). The dashed lines correspond to the Chieffi yields with an IMF slope of -2.7.}
\label{just-fig3}
\end{figure}

The yields of O and Mg depend strongly on the high-mass slope of the IMF. Unfortunately, there are inconsistent SN2 yields published in the literature. This results in a degeneracy of the IMF slope and the yield set in the [O/Mg] abundance ratio. An example, with the gas infall fixed to the fiducial case, is shown in Fig.~\ref{just-fig3} with the empirically calibrated yields of Fran\c{c}ois et al. (2004) and the theoretical yields of Chieffi \& Limongi (2004). The Chieffi yields combined with an IMF slope of -2.7 (dashed lines) lead to the same [O/Mg]$\approx$0.12\,dex (vertical offset in Fig.~\ref{just-fig3}) as the Fran\c{c}ois yields with an IMF slope of -2.3. On the other hand the oxygen yields depend much stronger on the IMF slope than the Mg yields. We conclude that it is required for a consistent disc model to have a more detailed look on the different $\alpha$-elements (and clearly define, which elements are used to determine the $\alpha$-enhancement) in order to fix the high-mass IMF slope and to determine the correct yields.

\begin{figure}
\includegraphics[width=0.48\textwidth]{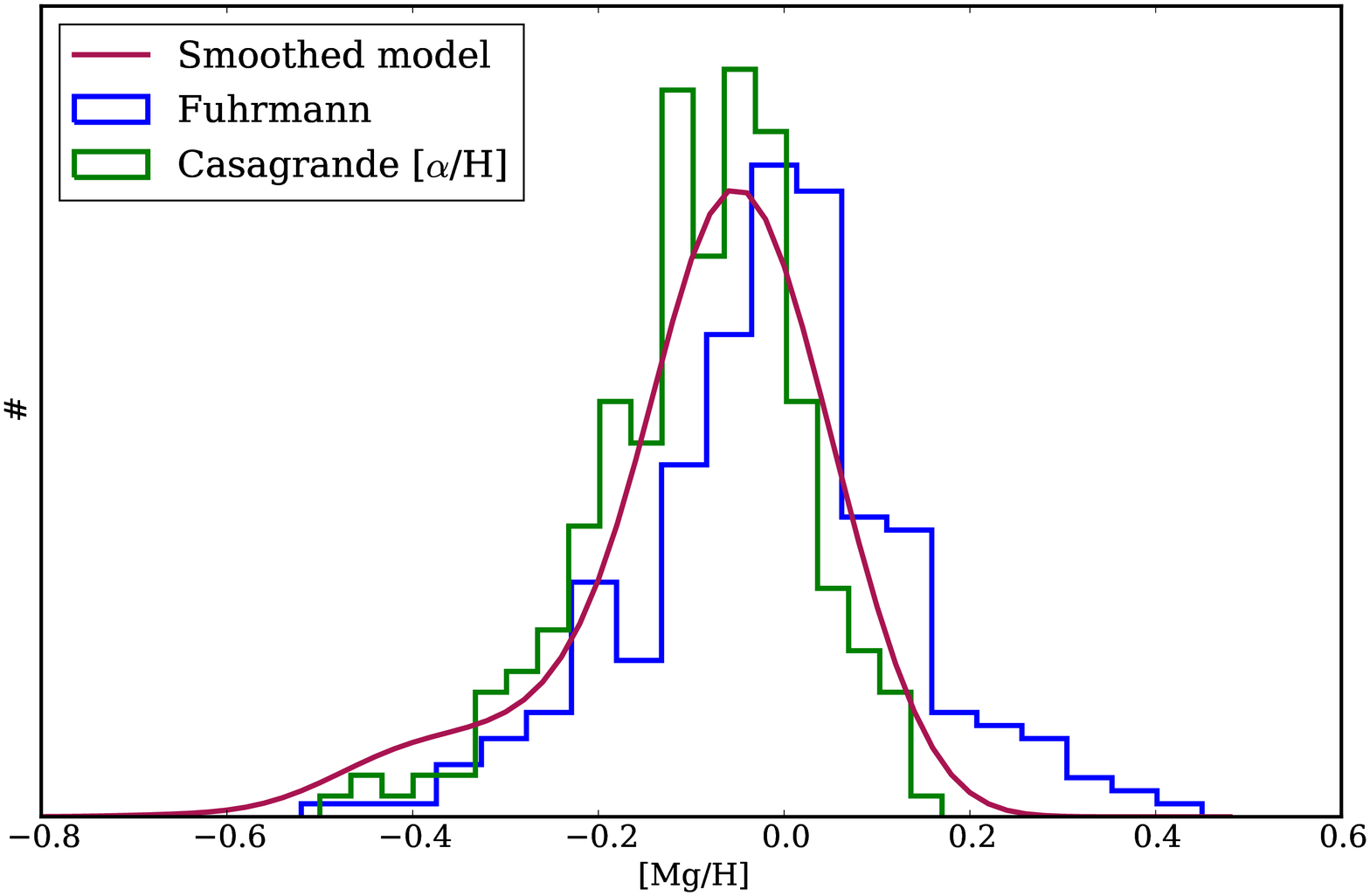} %{sfr_weighted_Mg_distribution_for_ms-stars}
\includegraphics[width=0.48\textwidth]{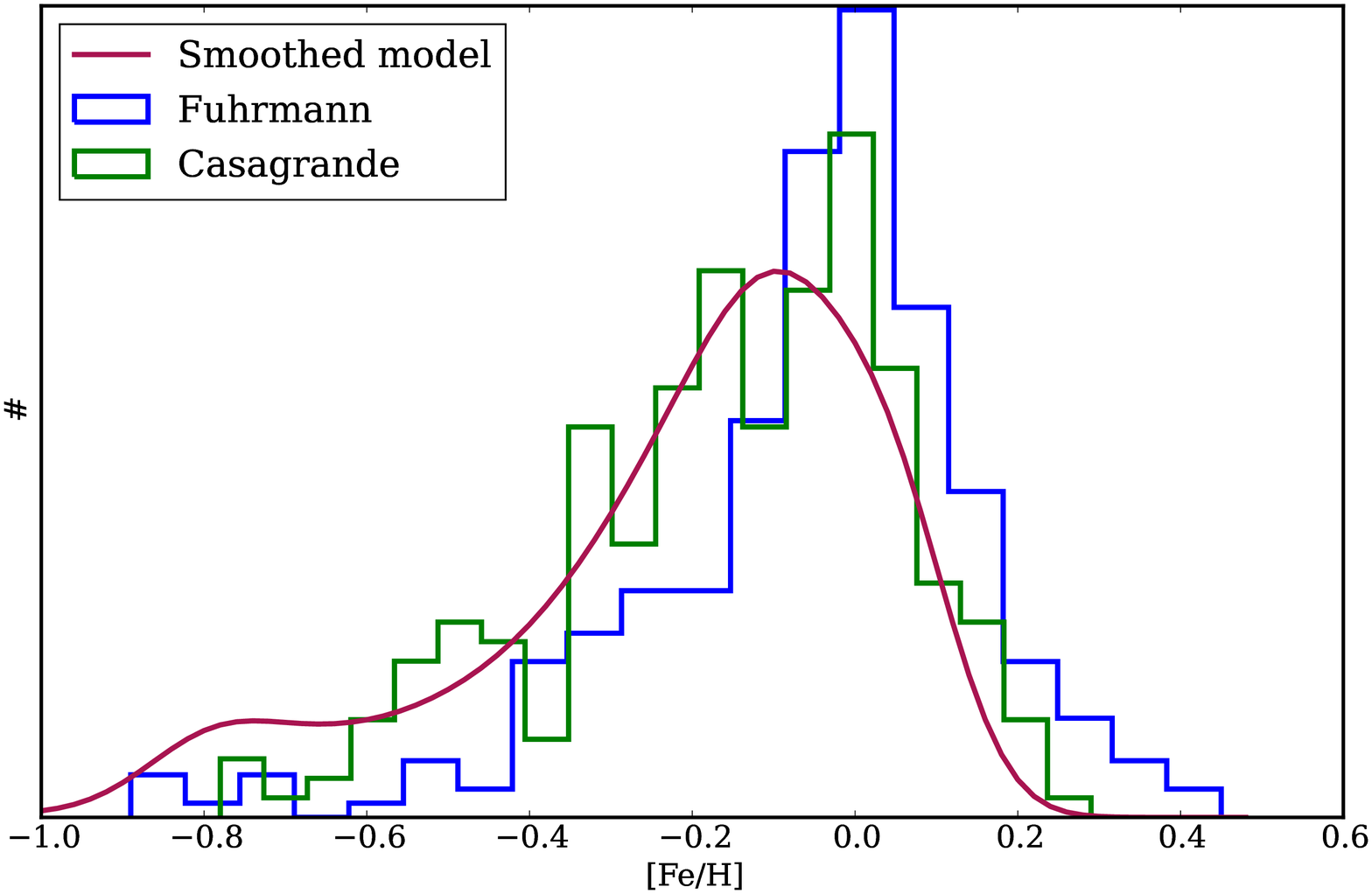} %{sfr_weighted_iron_distribution_for_ms-stars}
\caption{Normalised [Mg/H] and [Fe/H] distributions compared to the datasets of Casagrande and Fuhrmann.}
\label{just-fig4}
\end{figure}

After the determination of the gas infall rate and the IMF slope based on the $\alpha$-element distribution, we derive in the next step the distribution in [Fe/H]. The yields of supernovae type 1a (SN1a) are usually parametrised by the delay time distribution (DTD) and scaled by a number fraction of planetary nebulae (PN) exploding as SN1a. We have chosen a DTD with maximum at 1\,Gyr and a decay timescale of 2.5\,Gyr combined with a fraction of 0.2\% exploding PNs. The resulting abundance distributions are smoothed by a rms scatter of 0.07\,dex. In Fig.~\ref{just-fig4} the [Mg/H] and [Fe/H] distributions are shown in comparison to the Casagrande and the Fuhrmann sample. There are still some issues to be solved (the low metallicity bump and the systematic shift at the high metallicity end), but the general element distributions and correlations can be reproduced in the framework of the local JJ-model.

\section{Age distributions and vertical gradients}

\begin{figure}
\includegraphics[width=0.48\textwidth]{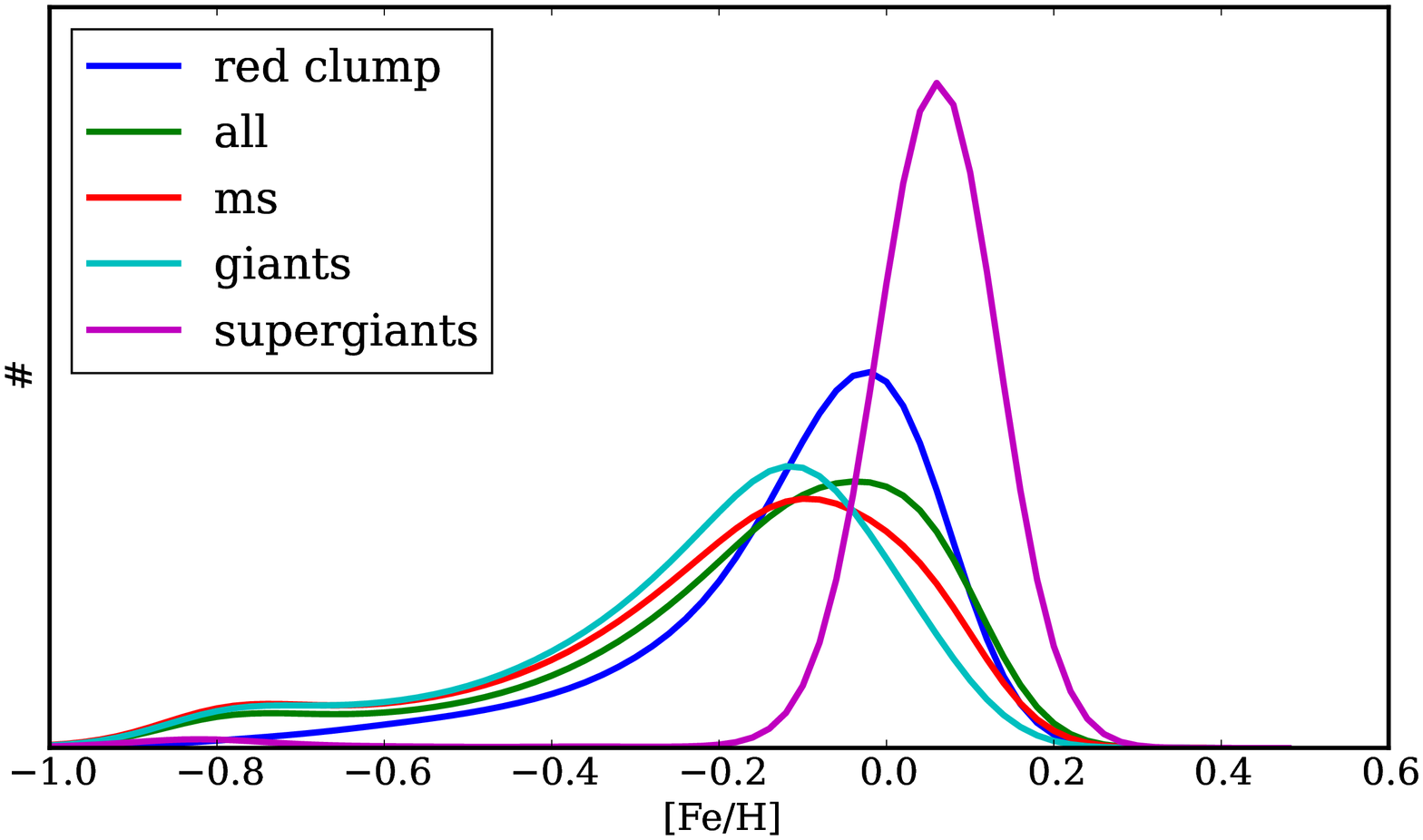} %{4_1}
\includegraphics[width=0.48\textwidth]{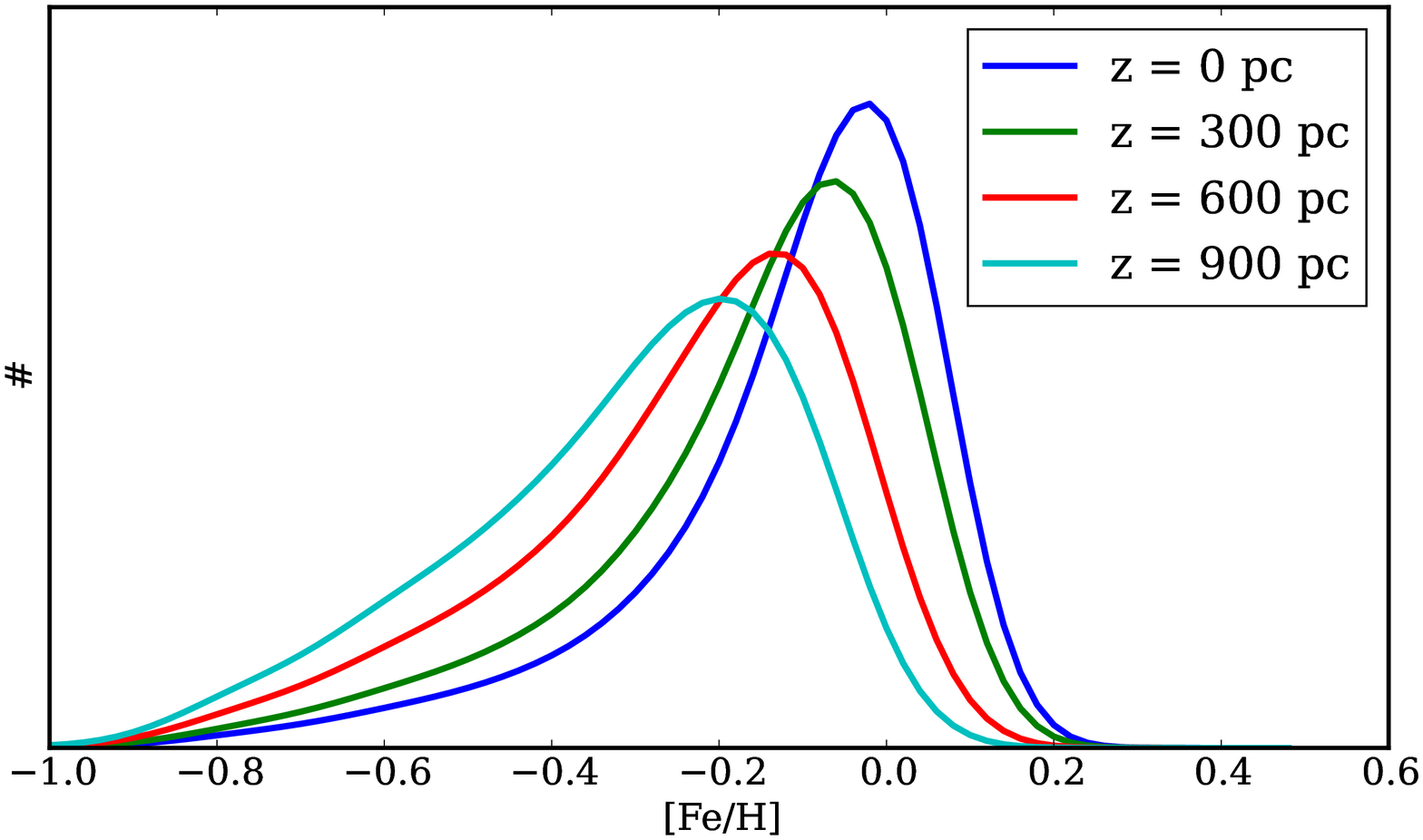} %{4_2}
\includegraphics[width=0.48\textwidth]{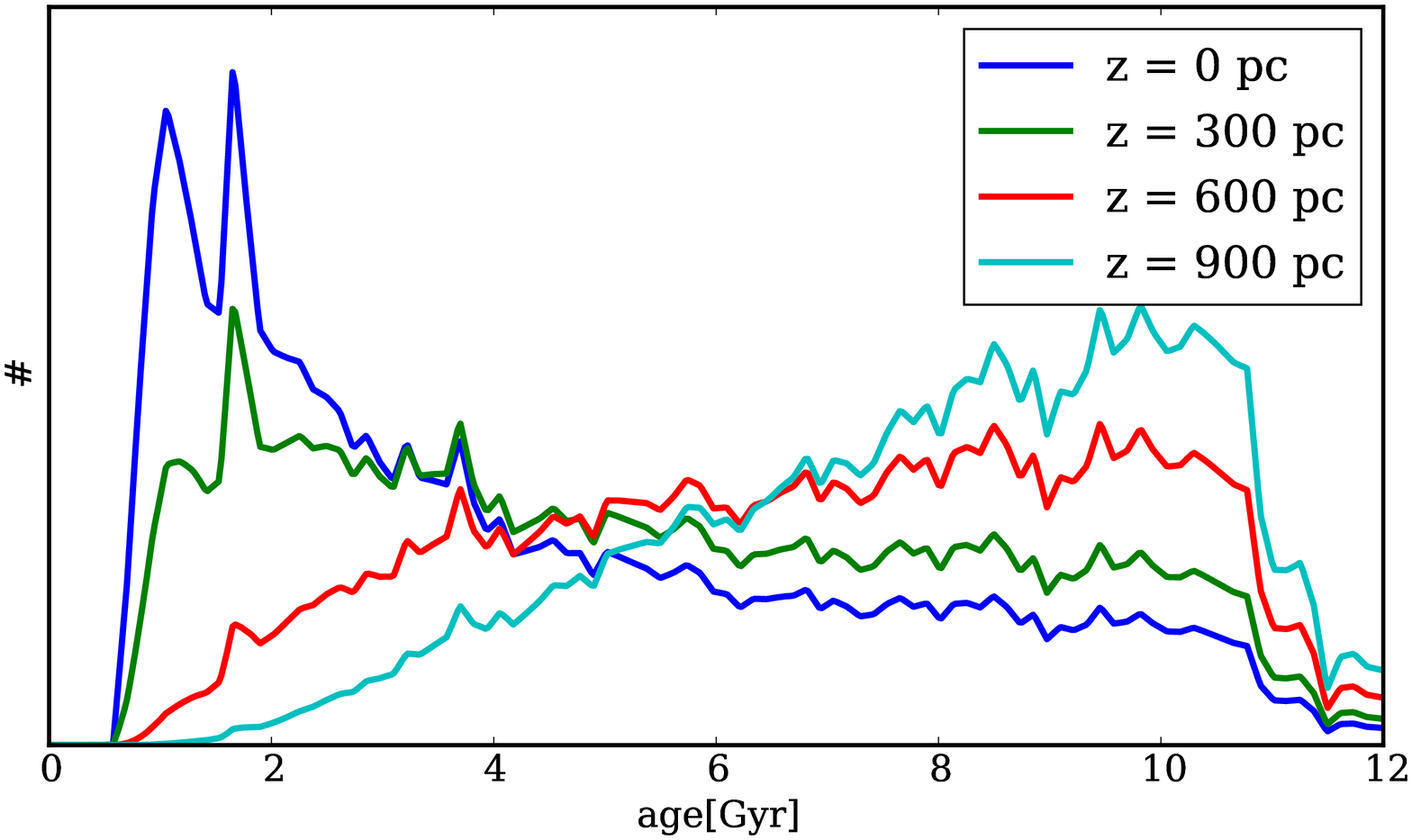} %{4_3}
\caption{Top panel: [Fe/H] distributions of different stellar types in the solar neighbourhood as predicted by the JJ-model.
Middle panel: Predicted [Fe/H] distributions of red clump stars with increasing $|z|$ above the Galactic plane.
Bottom panel: Age distributions of red clump stars at different $|z|$.}
\label{just-fig5}
\end{figure}

Since different types of stars have different age distributions due to different evolutionary stages, the resulting [Fe/H] abundance distributions do also vary. The resulting [Fe/H] distributions in the solar neighbourhood for lower main sequence, red clump, giant, and supergiant stars are shown in the top panel of Fig.~\ref{just-fig5} as predicted by the JJ-model. Especially the red clump stars are very important, because they can be identified easily in the CMD and observed over large distances.
The dynamical evolution of the thin disc results in vertical gradients of the stellar age distribution and the corresponding abundance distributions. The middle panel of Fig.~\ref{just-fig5} shows the strong variation of the [Fe/H] distribution of red clump stars with distance $|z|$ from the Galactic plane. The bottom panel of Fig.~\ref{just-fig5} shows the corresponding age distributions. This demonstrates that the intrinsic structure of the thin disc alone leads to strong vertical gradients, e.g. a significant shift of the maximum by -0.2\,dex at 900\,pc. The increasing contribution of the thick disc with increasing distance to the mid-plane enhances the gradients additionally (but is not included here).

\section{Radial gradients and inside-out growth}

\begin{figure}
\includegraphics[width=0.48\textwidth]{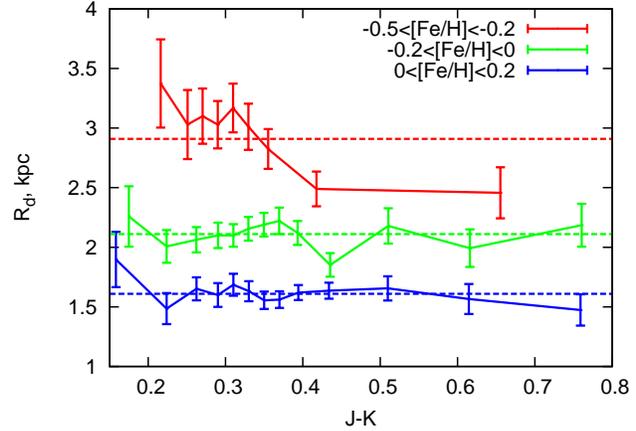}  %{Golubov13-fig7}
\caption{Radial scale-lengths at different metallicity bins of RAVE data as determined by the Jeans analysis of the asymmetric drift (from Golubov et al. 2013). Data points are colour bins along the main sequence.}
\label{just-fig6}
\end{figure}

The extension of the chemo-dynamical disc model over the full radial distance range depends on a full 3-dimensional model and requires homogeneous datasets covering a large range of Galactocentric distances. With a Jeans analysis of the asymmetric drift of RAVE data we have shown that the radial scale-length depends on metallicity (Golubov et al. 2013). The radial scale-length decreases from 2.9\,kpc at low metallicity to 1.6\,kpc at super-solar metallicity  and is essentially independent of colour along the main sequence (see Fig.~\ref{just-fig6}). Only in the low metallicity bin there is a decline with colour, which may originate from the contribution of the thick disc. A consistent extrapolation of these radial scale-lengths in the framework of a disc formation and enrichment model, where the SFR, AVR, and gas infall depend on Galactocentric distance, can be tested and constraint by using the recent spectroscopic surveys RAVE (Kordopatis et al. 2013), SDSS/APOGEE (Hayden et al. 2014, Holtzman et al. 2015), and
Gaia-ESO (GES, Smiljanic et al. 2014; Mikolaitis et al. 2014). Radial and vertical metallicity gradients are well identified in these surveys and the meridional plane distribution of the $\alpha$-enhancement and other abundance ratios can be derived. 
Ultimately, adding the high-precision parallaxes and proper motions expected from the first full data release of the Gaia mission in summer 2017 will allow direct measurements of density profiles and kinematic properties over a couple of kpc in the disc.
Based on these data we expect to be able to construct a self-consistent evolutionary thin disc model, which allows to determine the inside-out growth of the disc. An important task will also be to disentangle the thin and thick discs, and to quantify the impact of radial migration necessary to understand the chemical and dynamical properties of the stellar disc(s).

\acknowledgements
This work was supported by Sonderforschungsbereich SFB 881 "The Milky Way System" (sub-project A6) of the German Research Foundation (DFG). The work contains contributions of the undergraduate students Sarah Casura and Simon Sauer.


\begin{thebibliography}{}
\bibitem[Bressan et al. (2012)]{Bressan2012}
Bressan, A., Marigo, P., Girardi, L., et al. 2012, \mnras, 427, 127
\bibitem{}
Casagrande, L., Sch\"onrich, R., Asplund, M., et al. 2011, \aap, 530, A138
\bibitem{}
Chieffi, A., Limongi, M. 2004, \apj, 608, 405
\bibitem{}
Czekaj, M.A., Robin, A.C., Figueras, F., Luri, X., Haywood, M. 2014, \aap, 564, A102
\bibitem{}
Fran\c{c}ois, P., Matteucci, F., Cayrel, R., et al. 2004, \aap, 421, 613
\bibitem{}
Fuhrmann, K. 2011, \mnras, 414, 2893
\bibitem{}
Gao, S., Just, A., Grebel, E.K. 2013, \aap, 549, A20
\bibitem{}
Golubov. O., Just, A., Bienaym\'e, O., et al. 2013, \aap, 557, A92
\bibitem{}
Hayden, M.R., Holtzman, J.A., Bovy, J., et al. 2014, \aj, 147, 116
\bibitem{}
Holtzman, J.A., Shetrone, M., Johnson, J.A., et al., subm. to AJ, (astroph 1501.04110)
\bibitem{}
Just, A., Gao, S., Vidrih, S. 2011, \mnras, 411, 2586
\bibitem{}
Just, A., Jahrei\ss, H. 2010, \mnras, 402, 461
\bibitem{}
Kordopatis, G., Gilmore, G., Steinmetz, M., et al. 2013, \apj, 146, 134
\bibitem{}
Kubryk, M., Prantzos, N., Athanassoula, E. 2015a,b, \aap, 570, A126, A127
\bibitem{}
Lee, S.L., Beers, T.C., An, D., et al. 2011, \apj, 738, 137
\bibitem{}
Matteucci, F. 2012, Chemical Evolution of Galaxies. Springer-Verlag, Berlin, Heidelberg
\bibitem{}
Mikolaitis, \u{S}., Hill, V., Recio-Blanco, A., et al. 2014, \aap, 572, A33
\bibitem{}
Minchev, I., Chiappini, C., Martig, M. 2014, \aap 572, A92
\bibitem{}
Rybizki, J., Just, A. 2015 \mnras, 447, 3880
\bibitem{}
Sch\"onrich, R., Binney, J. 2009, \mnras, 399, 1145
\bibitem{}
Sharma, S., Bland-Hawthorn, J., Johnston, K.V., Binney, J. 2011, \apj, 730, 3
\bibitem{}
Smiljanic, R., Korn, A.J., Bergemann, M., et al. 2014, \aap, 570, A122
\end{thebibliography}
\end{document}